\documentclass[10pt,aps,prl,twocolumn,groupedaddress,showkeys]{revtex4-2}
\usepackage{amsmath}
\usepackage{graphicx}
\DeclareGraphicsExtensions{.pdf,.eps,.png,.jpg,.mps}
\usepackage{epstopdf}
\usepackage{threeparttable}
\usepackage{color}
\begin{document}
\title{Monolayer WS$_2$ electro- and photo-luminescence enhancement by TFSI treatment}
\author{A. R. Cadore$^1$}
\thanks{Present address: Brazilian Nanotechnology National Laboratory, Brazilian Center for Research in Energy and Materials, São Paulo, Brazil}
\author{B. L. T. Rosa$^1$}
\author{I. Paradisanos$^1$}
\author{S. Mignuzzi$^1$}
\author{D. De Fazio$^1$}
\author{E. M. Alexeev$^1$}
\author{J. E. Muench$^1$}
\author{G. Kakavelakis$^1$}
\author{S. M. Shinde$^1$}
\author{D. Yoon$^1$}
\author{S. Tongay$^2$}
\author{K. Watanabe$^3$}
\author{T. Taniguchi$^3$}
\author{E. Lidorikis$^4$}
\author{I. Goykhman$^1$}
\thanks{Present Address: Micro- and Nanoelectronics Research Center, Technion, Haifa, Israel}
\author{G. Soavi$^1$}
\thanks{Present Address: Institute of Solid State Physics, Friedrich Schiller University Jena, MAx-Wien Platz 1 Jena 07743, Germany and Abbe Center of Photonics, Friedrich Schiller University Jena, Albert-Einstein-Straße 6 Jena 07745, Germany}
\author{A. C. Ferrari$^1$}
\email{acf26@eng.cam.ac.uk}
\affiliation{$^1$Cambridge Graphene Centre, University of Cambridge, Cambridge CB3 0FA, UK}
\affiliation{$^2$School for Engineering of Matter, Transport and Energy, Arizona State University, Tempe, USA}
\affiliation{$^3$Advanced Materials Laboratory, 1-1 Namiki, Tsukuba, Japan}
\affiliation{$^4$Department of Materials Science and Engineering, University of Ioannina, Ioannina, Greece}
\begin{abstract}
Layered material heterostructures (LMHs) can be used to fabricate electroluminescent devices operating in the visible spectral region. A major advantage of LMH-light emitting diodes (LEDs) is that electroluminescence (EL) emission can be tuned across that of different exciton complexes (e.g. biexcitons, trions, quintons) by controlling the charge density. However, these devices have an EL quantum efficiency as low as$\sim$10$^{-4}$\%. Here, we show that the superacid bis-(triuoromethane)sulfonimide (TFSI) treatment of monolayer WS$_2$-LEDs boosts EL quantum efficiency by over one order of magnitude at room temperature. Non-treated devices emit light mainly from negatively charged excitons, while the emission in treated ones predominantly involves radiative recombination of neutral excitons. This paves the way to tunable and efficient LMH-LEDs.
\end{abstract}
\maketitle
Transition metal dichalcogenide monolayers (1L-TMDs) are ideal to study light-matter interactions and many-body effects at the atomic scale\cite{SchneiderNC2018,GWangPhysMod2018,MuellerNPJ2018}. Compared to bulk semiconductors\cite{GWangPhysMod2018}, the reduced dielectric screening combined with the spatial confinement of charge carriers\cite{SchneiderNC2018} favours the formation of various excitonic complexes which can be controlled by modulation of the carrier density\cite{SchneiderNC2018,GWangPhysMod2018,MuellerNPJ2018,Glazov2020,BarboneNC2018,WangNL2019,PaurNC2019,LienSci2019}. Thus, 1L-TMDs photoluminescence (PL) spectra host features arising from formation of charged\cite{Glazov2020,BarboneNC2018,WangNL2019,PaurNC2019,LienSci2019} and neutral\cite{ZhuScRep2015,ShangACSNano2015,PeimyooNL2014,YangNL2016} exciton complexes.

Layered material heterostructures (LMHs) combining single layer graphene (SLG), 1L-TMDs, and hexagonal boron nitride (hBN), from 1L-hBN to hundreds of layers, are promising for electronics\cite{Nanoscale2015,ManzeliNRM2017}, photonics\cite{Bonaccorso2010}, and optoelectronics\cite{Koppens2014,RomaNRM2018}. Direct bandgap 1L-TMDs and LMHs can be used to make light-emitting diodes (LEDs)\cite{ChoiMT2017,WangCSR2018,ZhengAOM2018,CarmenNC2016,JoNL2014,WithersNM2015,WithersNL2015,LiuNL2017,RomanNanoscale2020,WangNN2012}, with fast modulation speed (up to GHz)\cite{PaurNC2019,LiuNL2017,KwakADM21}, and emission wavelength tunability\cite{PaurNC2019,LiuNL2017,WangNL2019} besides multi-spectral (visible$\sim$618nm\cite{JoNL2014,CarmenNC2016,WithersNM2015} to near-infrared$\sim$1160nm\cite{ZhuACS2018,BieNN2017}) emission.

In 1L-TMD-based LEDs, the electroluminescence (EL) efficiency ($\eta_{EL}$), i.e. ratio between emitted photons and injected electrons (\textit{e})\cite{WangCSR2018,ZhengAOM2018}, depends on the optical emission of the material\cite{SundaranNL2013,ShengACS2019,AndrzejewskiACSPhot2019,AndrzejewskiNanoscale2019,GuNN2019,BieNN2017,RossNN2014,BaugherNN2014}, as well as on its doping level\cite{WangNL2017,WangNL2019,YuanNano2015,WangNL2015,ZhangSci2014}. In doped 1L-TMDs, the PL and EL emission originates from either negative (X$^{-}$)\cite{KwakADM21,WangNL2017,AndrzejewskiACSPhot2019,AndrzejewskiNanoscale2019} or positive (X$^{+}$)\cite{WangNL2019,WangCSR2018,ZhengAOM2018} trions, depending on the type of doping. However, 1L-TMD-LEDs based on trionic emission show low $\eta_{EL}$ (typically$<$0.05\%\cite{WangCSR2018,ZhengAOM2018}) with respect to neutral exciton (X$^{0}$) emission (typically $\eta_{EL}<$1\%\cite{SundaranNL2013,ShengACS2019,PaurNC2019,WangNL2017,WangNL2019,YuanNano2015}). This difference in $\eta_{EL}$ occurs due the small ($\sim$30meV) binding energy of trions\cite{JonesNatP2016}. Since the X$^{-}$ binding energy is close to the lattice thermal energy at room-temperature (RT=300K,$\sim$25.2meV), trions dissociate\cite{GWangPhysMod2018}. An excess of free-carriers decreases the available phase-space filling for exciton complexes, due to Pauli blocking, with a reduction of trion and exciton binding energies\cite{LundtAPL2018} and oscillator strengths\cite{HaugBook2009} (i.e. the probability of absorption/emission of electromagnetic radiation\cite{KlingshirnBook2007}).

In 1L-TMDs, low light-emission efficiency is observed in both EL ($\eta_{EL}\sim$10$^{-4}$\cite{AndrzejewskiACSPhot2019,AndrzejewskiNanoscale2019} to$\sim$1\%\cite{SundaranNL2013,ShengACS2019,PaurNC2019,WangNL2017,WangNL2019,YuanNano2015}) and PL ($\eta_{PL}$ $\sim$10$^{-3}$\cite{RossNN2014,WangNL2015} to$\sim$5\%\cite{SchneiderNC2018,GWangPhysMod2018,MuellerNPJ2018}). $\eta_{PL}$ is defined as the ratio between emitted and absorbed photons\cite{WangCSR2018,ZhengAOM2018}. Thus, several chemical approaches were suggested to enhance $\eta_{PL}$, such as treatment with 2,3,5,6-tetrafluoro 7,7,8,8-tetracyanoquinodimethane\cite{MouriNL2013}, hydrogen peroxide\cite{SuRSC2015}, titanyl phthalocyanine\cite{ParkScAd2017}, sulfuric acid\cite{KiriyaLang2018}, oleic acid\cite{TanohNL2019,BrestcherACS2021,Tanoh2021}, and the superacid (i.e. with acidity greater than that of 100\% pure sulfuric acid\cite{Acid-def}) bis-(trifluoromethane)sulfonimide (TFSI)\cite{AmaniSci2015,AmaniACS2016,AmaniNL2016,KimACS2016,MolasSRep2019,RoyNL2019,AlharbiAPL2017,CadizAPL2016,ZhangSRep2017,LinJMCC2019,GoodmanPRB2017,LuAPL2018,DhakalJMCC2017}. TFSI treatment increased the PL intensity of 1L-WS$_2$ up to$\sim$10-times\cite{AmaniSci2015,AmaniNL2016,AmaniACS2016,TanohNL2019} due to depletion of excess \textit{e}, promoting X$^{0}$ recombination.

The effect of chemical passivation of 1L-TMDs on $\eta_{EL}$ combined with gated-PL emission in 1L-TMD-based LEDs was not reported to date, to the best of our knowledge. Refs.\cite{AmaniSci2015,AmaniACS2016,AmaniNL2016,KimACS2016,MolasSRep2019,RoyNL2019,AlharbiAPL2017,CadizAPL2016} reported PL measurements on 1L-TMDs and focused on non-gated samples, thus limiting the modulation of charge density in 1L-TMDs. Ref.\cite{LienSci2019} performed gated-PL measurements in 1L-WS$_2$, finding that both TFSI treatment and electrical gating increase $\eta_{PL}$ by a factor of up to$\sim$10 (at$\sim$10$^{19}$cm$^{-2}$s$^{-1}$ photocarrier generation rate), because both processes reduce the \textit{n}-type behaviour of 1L-WS$_2$ and suppress X$^{-}$ formation, thus enhancing X$^{0}$ radiative recombination. However, gated-PL measurements after TFSI passivation were not provided. The activation of trapping states on TFSI-treated 1L-TMDs was not discussed. Ref.\cite{LienNC2018} carried out EL experiments with TFSI passivation for high-speed (MHz) modulation, but did not report PL nor EL emission tunability. Therefore, an investigation on how TFSI affects EL emission and modifies gated-PL of 1L-TMD-based devices is required.

Here, we fabricate LEDs with 1L-WS$_2$ as active material on a metal-insulator-semiconductor (MIS) structure. We measure EL and gated-PL before and after TFSI treatment. We find that TFSI increases $\eta_{EL}$ by over one order of magnitude at RT, and PL intensity by a factor$\sim$5. We find that X$^{-}$ and X$^{0}$ are present in both EL and PL before TFSI treatment, whereas X$^{0}$ dominates after. We attribute this to depletion of excess \textit{e} and changes in the relaxation pathway, induced by the treatment. This paves the way to more efficient 1L-TMDs-based LEDs and excitonic devices.
\section{Results and Discussion}
\begin{figure}
\centerline{\includegraphics[width=85mm]{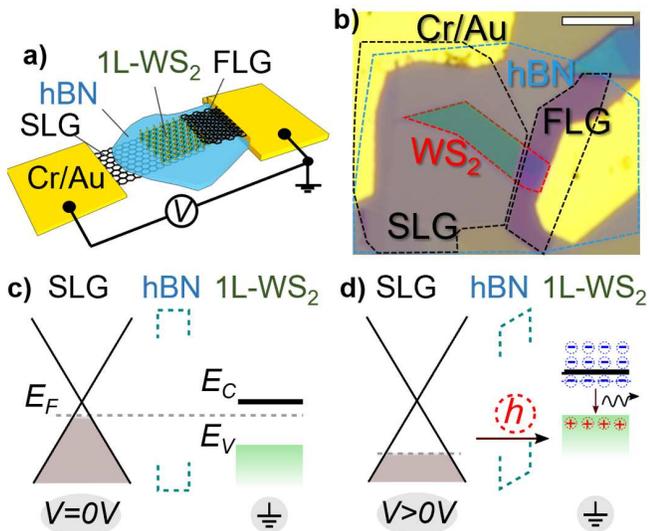}}
\caption{a) Schematic of LED. Cr/Au electrodes, SLG, FLG, hBN, and 1L-WS$_2$ are indicated. b) Optical image of device. Scale bar 4$\mu$m. The dotted lines highlight the footprint of SLG, FLG, hBN, 1L-WS$_2$. The green-shaded part corresponds to the active area$\sim$23$\mu$m$^{2}$. Cr/Au contacts the bottom SLG; FLG contacts the top 1L-WS$_2$. Band diagram for (c)\emph{V}=0V and (d) \emph{V}$>$0V. Tuning the SLG E$_F$ (gray dotted line) across the 1L-WS$_2$ valence band edge, E$_V$, allows \textit{h} tunneling from SLG to 1L-WS$_2$, resulting in current onset and light emission via radiative recombination with \textit{e} from the \textit{n}-type 1L-WS$_2$. The blue circles represent \textit{e} accumulated on 1L-WS$_2$ due to the MIS structure, while the red circles are \textit{h} injected into 1L-WS$_2$ through the hBN barrier}
\label{fig:Fig1}
\end{figure}
We use 1L-WS$_2$ as the active light-emitting layer since it has a direct bandgap\cite{BerSRep2013,GutNL2013,Molina-SanchezPRB2011,ZhaoNanoscale2013}, its PL emission is$\sim$60 times stronger than 1L-MoS$_2$\cite{GutNL2013,YuanNano2015} at RT, $\eta_{EL}$ can be up to$\sim$50 times larger than 1L-MoS$_2$\cite{WangCSR2018,ZhengAOM2018} at RT, while Refs.\cite{TanohNL2019,AmaniSci2015,AmaniACS2016,AmaniNL2016,KimACS2016,MolasSRep2019,RoyNL2019,AlharbiAPL2017,CadizAPL2016,ZhangSRep2017,LinJMCC2019,GoodmanPRB2017,LuAPL2018} demonstrated that TFSI treatment increases up to$\sim$10-times its PL intensity.

Fig.\ref{fig:Fig1}a shows the 1L-WS$_2$/hBN/SLG tunnel junction configuration used here, where the metallic electrodes provide contacts to apply a voltage (\emph{V}) between SLG and 1L-WS$_2$. This is prepared as follows.

WS$_2$ crystals are synthesized using a two-step self-flux technique\cite{Nagao2017} using 99.9999\% purity W and S powders without any transporting agents. Commercial (Alfa Aesar) sources of powders contain a number of defects and impurities (Li, O, Na, and other metals as determined by secondary ion mass spectroscopy). Before growth, W and S powders are thus purified using electrolytic\cite{Suchkov1971} and H$_2$\cite{Suchkov1971} based techniques to reach 99.995\% purity. WS$_2$ polycrystalline powders are created by annealing a stoichiometric ratio of powders at 900$^{\circ}$C for 3 weeks in a quartz ampoule sealed at 10$^{-7}$ Torr. The resulting powders are re-sealed in a different quartz ampoule under similar pressures and further annealed at 870-910$^{\circ}$C with thermodynamic temperature differential (hot to cold zone difference)$\sim$40$^{\circ}$C. The growth process takes 5 weeks. At the end of the growth, ampoules are cooled to RT slowly ($\sim$40$^{\circ}$C/hour)\cite{Montblanch2021}. We use this material as bulk source because our previous work\cite{Montblanch2021} demonstrated that this has a point defect density$\sim$10$^{9}$-10$^{10}$ cm$^{-2}$, on par or better than previous reports\cite{EdelbergNL2019}.

Bulk WS$_2$, hBN (grown by the temperature-gradient method\cite{WatNM2004}), and graphite (sourced from HQ Graphene) crystals are then exfoliated by micromechanical cleavage using Nitto-tape\cite{Novoselov2005} on 285nm SiO$_2$/Si. Optical contrast\cite{CasiNL2007} is first used to identify 1L-WS$_2$, SLG, FLG (3-10nm), and hBN($<$5nm). The LMs are then characterized by Raman spectroscopy as discussed in Methods. After Raman characterization of all individual LMs on SiO$_2$/Si, the FLG/1L-WS$_2$/hBN/SLG LMH is assembled using dry-transfer as for Refs.\cite{Purdie2018,Viti2021}. FLG is picked-up from SiO$_2$/Si using a polycarbonate (PC) membrane on a polydimethylsiloxane (PDMS) stamp (as mechanical support) at 40$^{\circ}$C. We use 40$^{\circ}$C because this is sufficient to increase the adhesion of the PC film\cite{PizzoccheroNC16}, to pick all LMs from SiO$_2$/Si. Then, FLG is aligned to one edge of 1L-WS$_2$ on SiO$_2$/Si and brought into contact using \textit{xyz} micromanipulators at 40$^{\circ}$C, leaving the majority of 1L-WS$_2$ without FLG cover to be used as active area (AA). AA is the region from where light emission is expected, and it is the overlap area between 1L-WS$_2$ and SLG (green-shaded part in Fig.\ref{fig:Fig1}b). Next, FLG/1L-WS$_2$ is aligned to a hBN flake deposited onto SiO$_2$/Si and brought into contact using \textit{xyz} micromanipulators at 40$^{\circ}$C. Finally, FLG/1L-WS$_2$/hBN is aligned to a SLG on SiO$_2$/Si and brought into contact using \textit{xyz} micromanipulators at 180$^{\circ}$C, whereby PC preferentially adheres to SiO$_2$\cite{Purdie2018}, allowing PDMS to be peeled away, leaving PC/FLG/1L-WS$_2$/hBN/SLG on SiO$_2$/Si. PC is then dissolved in chloroform for$\sim$15mins at RT, leaving the FLG/1L-WS$_2$/hBN/SLG LMH on SiO$_2$/Si\cite{Purdie2018,Viti2021}. After LMH assembly, Cr/Au electrodes are fabricated by electron beam lithography (EBPG 5200, Raith GMBH), followed by metallization (1:50nm) and lift-off.

The tunnel junction based on a MIS structure consists of a LMH with 1L-WS$_2$ as the light emitter, FL-hBN (typically from 2 to 4nm) acting as tunnel barrier, and a SLG electrode to inject holes (\textit{h}) into 1L-WS$_2$. We use FL-hBN$<$5nm so that a low (typically$<$5V) driving voltage is sufficient for charge injection to the 1L-WS$_2$\cite{HatACSN2015,Man2D2017}. We employ FLG ($\sim$3-10nm) to contact 1L-WS$_2$, because FLG reduces the contact resistance\cite{GuimaraesACS2016}, while Cr/Au electrodes give Ohmic contacts to SLG and FLG\cite{GuimaraesACS2016}. SLG could also be used to contact 1L-WS$_2$, however, as the optical contrast is higher in FLG than SLG\cite{NairScience2008,CasiNL2007}, using FLG makes it easier to align it to 1L-WS$_2$ during transfer. Since TFSI treatment requires direct exposure of 1L-TMDs\cite{AmaniSci2015}, we place 1L-WS$_2$ on top of the stack to compare the device performance before and after treatment. We TFSI-treat 4 samples for EL and gated-PL measurements. These are immersed in a TFSI solution (0.2 mg/mL) in a closed vial for 10mins at 100$^{\circ}$C\cite{AmaniSci2015,AmaniACS2016,AmaniNL2016}, then removed, dried by a N$_2$ gun, and annealed on a hot plate at 100$^{\circ}$C for 5mins\cite{AmaniSci2015,AmaniACS2016,AmaniNL2016}. Fig.\ref{fig:Fig1}b is an image of the 1L-WS$_2$-LEDs. The FLG electrode is placed on the side of the SLG to avoid direct tunneling of carriers from SLG to FLG, hence keeping as AA the LMH region extended over SLG and 1L-WS$_2$, green-shaded in Fig.\ref{fig:Fig1}b. If there is a FLG/SLG overlap, tunneling through FLG-SLG may be possible, not resulting in \textit{e}-\textit{h} recombination into 1L-WS$_2$, hence no EL\cite{LiuNL2017,WangNL2017,WangNL2019}.

Figs.\ref{fig:Fig1}c,d sketch the band diagram of our LEDs for \emph{V}=0V and \emph{V}$>$0V, respectively. For \emph{V}=0V (at thermodynamic equilibrium as indicated in Fig.\ref{fig:Fig1}c), the Fermi level, E$_F$, is constant across the junction, and the net current (\emph{I}) is zero\cite{CarmenNC2016,LiuNL2017,WangNL2017,WangNL2019,KwakADM21}. For \emph{V}$>$0V (positive potential on SLG), the SLG E$_F$ is shifted below the 1L-WS$_2$ valence band energy E$_V$ (Fig.\ref{fig:Fig1}d), and \textit{h} from SLG tunnel across the hBN barrier into 1L-WS$_2$, promoting EL emission by radiative recombination between the injected excess \textit{h} and intrinsic \textit{e}\cite{CarmenNC2016,JoNL2014,WithersNM2015,WithersNL2015,WangNL2017,GuNN2019,KwakADM21}. The EL emission is expected to increase as a function of tunneling current because of the increasing \textit{h} injected into 1L-WS$_2$ available for \textit{e}-\textit{h} recombination.

The LMs are characterized by Raman, PL, EL spectroscopy using a Horiba LabRam HR Evolution. The Raman spectra are collected using a 100x objective with numerical aperture (NA)=0.9, and a 514.5nm laser with a power$\sim$5$\mu$W to avoid damage or heating. The voltage bias dependent PL and EL are collected using a long working distance 50x objective (NA=0.45). For the PL spectra, we use a 532nm (2.33eV) laser in order to excite above the X$^{0}$ emission ($\sim$2eV)\cite{ZhuScRep2015,ShangACSNano2015}. The power is kept$\sim$80nW to avoid laser-induced thermal effects\cite{GWangPhysMod2018,ZhuScRep2015,ShangACSNano2015,PeimyooNL2014}. The voltage (\emph{V}) and current (\emph{I}) between source (SLG) and drain (1L-WS$_2$) electrodes are set (\emph{V}) and measured (\emph{I}) by a Keithley 2400.
\begin{figure}
\centerline{\includegraphics[width=90mm]{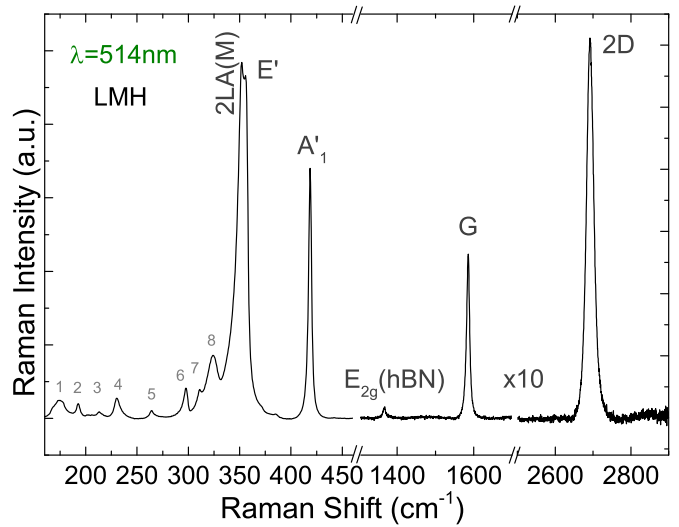}}
\caption{514.5nm Raman spectrum of 1L-WS$_2$/hBN/SLG LMH after device fabrication. The SLG and hBN Raman modes are labelled on it and the modes for 1L-WS$_2$ as for Table 1. The 1300-2900cm$^{-1}$ spectral window was multiplied by a factor of 10 for better visualization}
\label{fig:Fig2}
\end{figure}
\begin{figure}
\centerline{\includegraphics[width=90mm]{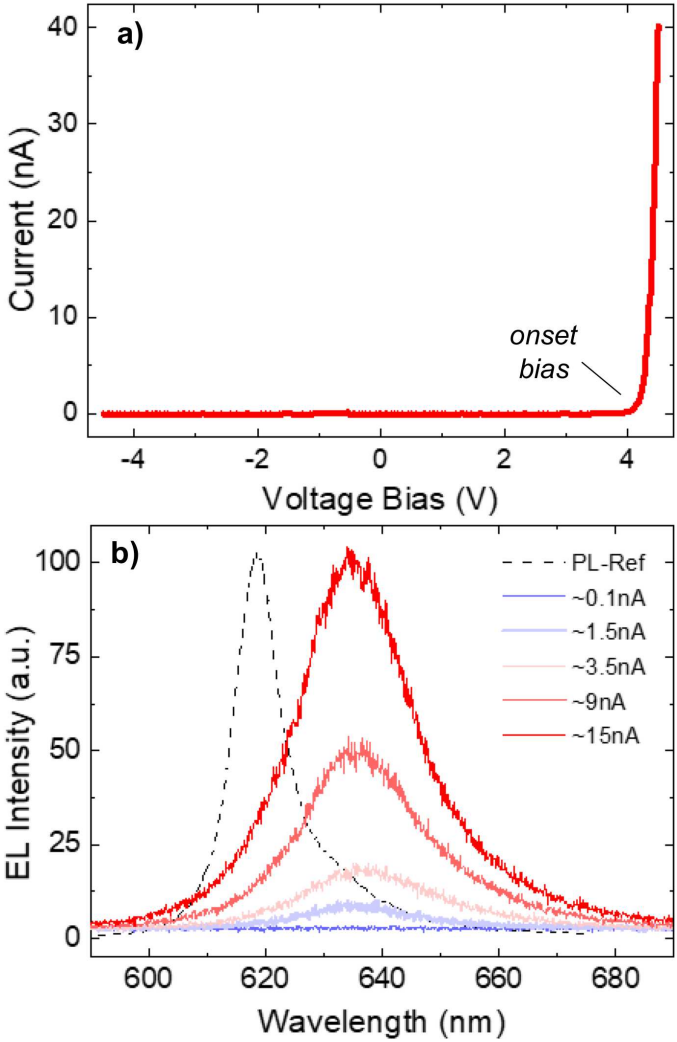}}
\caption{a) \emph{I} as a function of \emph{V} for 1L-WS$_2$-LED. b) EL spectra for different tunneling currents without TFSI treatment. The dashed black line is the PL spectrum collected at \emph{V}=0 and normalized to the maximum EL intensity.}
\label{fig:Fig3}
\end{figure}

Fig.\ref{fig:Fig2} shows the Raman spectrum of 1L-WS$_2$/hBN/SLG on Si/SiO$_2$ after device fabrication and before current-voltage (\emph{I-V}) measurements. The Raman modes of each LM can be identified. For 1L-WS$_2$, Pos(A$_{1}^{'}$) and its full width af half maximum, FWHM(A$_{1}^{'}$), change from$\sim$418.9$\pm0.2$cm$^{-1}$; 3.9$\pm0.2$cm$^{-1}$, before assembly, to$\sim$419.8$\pm0.2$cm$^{-1}$; 3.4$\pm0.2$cm$^{-1}$, after. All the changes in the other modes are close to our spectral resolution and errors, as for Ref.\cite{Keftoot2022}. Pos(A$_{1}^{'}$) and FWHM(A$_{1}^{'}$) are sensitive to changes in \textit{n}-doping\cite{ChakraborthyPRB2012,SohierPRX2019}. The mechanism responsible for this effect is an enhancement of electron-phonon (e-ph) coupling when \textit{e} populate the valleys at K and Q simultaneously\cite{SohierPRX2019}. The energy of the K and Q valleys is modulated by the A$_{1}^{'}$ ph\cite{SohierPRX2019}. Since the K and Q energies are modulated out-of-phase, charge transfer between the two valleys occurs in presence of the A$_{1}^{'}$ ph\cite{ChakraborthyPRB2012,SohierPRX2019}. When the K and Q valleys are populated by \textit{e}, these are transferred back and forward from one valley to the other\cite{SohierPRX2019,ParadNC2021}. This increases the e-ph coupling of out-of-plane modes, such as A$_{1}^{'}$\cite{SohierPRX2019}. The same process does not occur for \textit{p}-doping\cite{SohierPRX2019}. The reason for this asymmetry between \textit{n-} and \textit{p}-doping is due to a much larger energy separation ($\sim$230meV\cite{SohierPRX2019}) between the VB $\Gamma$ and K valleys than that ($\sim$100meV\cite{SohierPRX2019}) of the CB K and Q valleys. From the changes in Pos(A$_{1}^{'}$) and FWHM(A$_{1}^{'}$), and by comparison with Ref.\cite{SohierPRX2019}, we estimate a reduction in \textit{n}-doping$\sim5\times10^{12}$cm$^{-2}$.

For hBN in Fig.\ref{fig:Fig2}, Pos(E$_{2g}$)$\sim$1366.4$\pm0.2$cm$^{-1}$ and FWHM(E$_{2g}$)$\sim$9.2$\pm0.2$cm$^{-1}$. Although FWHM(E$_{2g}$) changes within the error, Pos(E$_{2g}$) downshifts$\sim$2.1cm$^{-1}$ after assembly, suggesting a contribution from strain (see Methods for comparison between FL- and bulk-hBN Raman). Uniaxial strain lifts the degeneracy of the E$_{2g}$ mode and results in the splitting in two subpeaks E$_{2g}^+$ and  E$_{2g}^-$, with shift rates$\sim$-8.4 and -25.2cm$^{-1}$/$\%$\cite{AndrPRB18,CaiNanos2017}. For small levels of uniaxial strain ($<$0.5$\%$) splitting cannot be observed and the shift rate is$\sim$-16.8cm$^{-1}/\%$\cite{AndrPRB18,CaiNanos2017}. For biaxial strain, splitting does not occur and E$_{2g}$ shifts with rate$\sim$-39.1cm$^{-1}$/$\%$\cite{AndrPRB18}. Since we do not observe splitting, the E$_{2g}$ shift can be attributed to uniaxial or biaxial tensile strain$\sim$0.13$\%$ or$\sim$0.06$\%$, respectively.

For SLG in Fig.\ref{fig:Fig2}, no D peak is observed after LMH assembly, indicating negligible defects\cite{Cancado2011,FerCom2007,FerNN2013}. In Fig.\ref{fig:Fig2} Pos(G)$\sim$1585.1$\pm0.2$cm$^{-1}$, FWHM(G)$\sim$9.0$\pm0.2$cm$^{-1}$, Pos(2D)$\sim$2692.3$\pm0.2$cm$^{-1}$, FWHM(2D)$\sim$20.9$\pm0.2$cm$^{-1}$, I(2D)/I(G)$\sim$2.4, and A(2D)/A(G)$\sim$5.6. These indicate that the SLG is \textit{p}-doped, with E$_F\sim$150$\pm$50meV\cite{Basko2009,FerCom2007,FerNN2013} by taking into account the average dielectric constant ($\sim$3.85) of the environment ($\varepsilon_{SiO_2}\sim$3.8\cite{KingonNature2000} and $\varepsilon_{hBN}\sim$3.9\cite{LaturiaNPJ2018}). E$_F\sim$150meV should correspond to Pos(G)$\sim$1584.1cm$^{-1}$ for unstrained SLG\cite{Das2008}. However, Pos(G)$\sim$1585.1$\pm0.2$cm$^{-1}$, which implies a contribution from compressive uniaxial (biaxial) strain$\sim$0.04$\%$ ($\sim$0.01$\%$). The strain level for SLG and hBN are different, most likely due to the fact that the SLG is directly exfoliated onto SiO$_2$/Si, while hBN is picked up and transferred by PDMS stamps, hence, this could induce a larger amount of strain on hBN.

Fig.\ref{fig:Fig3}a plots the \emph{I-V} characteristics. For \emph{V}=0V the current is zero (Fig.\ref{fig:Fig1}c). When \textit{V} is applied, an electrical rectification (i.e. diode behavior) with negligible leakage current (\emph{I}$<$10$^{-11}$A) for \emph{V}$<$0 is seen. A tunneling onset, (i.e. exponential increase of \emph{I}) is seen at \emph{V$_{ON}$}$\sim$4.1V, Fig.\ref{fig:Fig3}a. \emph{V$_{ON}$} is related to the breakdown electric field (E$_{bd}$) across the junction, which depends on the voltage drop on the hBN tunnel barrier and hBN thickness (\textit{d}) accordingly to E$_{bd}$=(V$_{bd}$/\textit{d})$\sim$0.7-1V/nm\cite{HatACSN2015,Man2D2017}, where V$_{bd}$ is voltage breakdown V$_{bd}$=\textit{qnd}$^2$/($\varepsilon_0\varepsilon_{hBN}$), \emph{q} is the \textit{e} charge, \textit{n} is total charge concentration, $\varepsilon_0$=8.854$\times$10$^{-12}$ F/m and $\varepsilon_{hBN}\sim$3.9\cite{HatACSN2015,Man2D2017}, so that \emph{V$_{ON}$} can vary between different devices. When \emph{V}$>$\emph{V}$_{ON}$, \textit{h} from SLG tunnel across the hBN barrier into 1L-WS$_2$, promoting EL emission by radiative recombination between the injected \textit{h} and majority \textit{e} in 1L-WS$_2$ (Fig.\ref{fig:Fig1}c)\cite{CarmenNC2016,JoNL2014,WithersNM2015,WithersNL2015,WangNL2017,GuNN2019}. The EL intensity$\sim$634nm ($\sim$1.956eV) increases with tunneling current, as in Fig.\ref{fig:Fig3}b. No light emission is observed in reverse \emph{V}$<0V$ and small positive ($0<$\emph{V}$<$\emph{V$_{ON}$}) biases, below the tunneling condition (\emph{V$_{ON}$}$<$4.1V). A red-shift$\sim$48meV is observed in EL emission$\sim$634nm ($\sim$1.956eV) with respect to the PL X$^0$ emission of the unbiased device (dashed black line, Fig.\ref{fig:Fig3}b). Fig.\ref{fig:Fig3}b shows a EL peak position close to X$^{-}$ of unbiased PL (dashed black line, Fig.\ref{fig:Fig3}b), implying a trionic EL emission, due to excess \textit{e} in 1L-WS$_2$\cite{WangNL2017,KwakADM21}.
\begin{figure}
\centerline{\includegraphics[width=90mm]{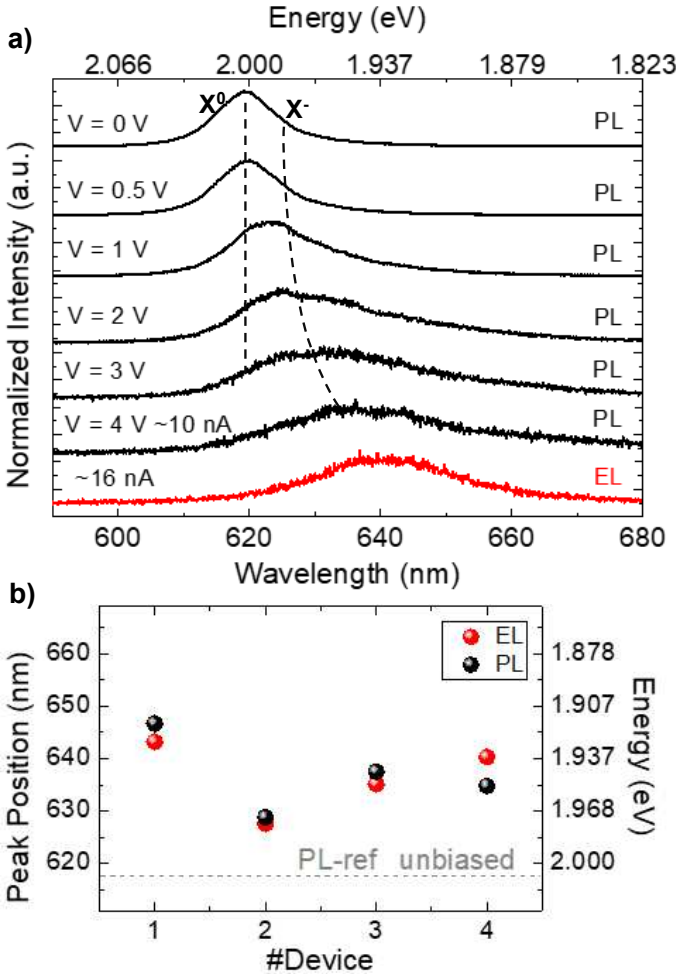}}
\caption{a) Evolution of PL as a function of \emph{V}. For comparison, an EL spectrum for I$\sim$16nA is shown (red). The dashed lines are guides to the eye for the X$^0$ and X$^{-}$ positions. In all PL measurements up to 3V, \emph{I}$<$10$^{-11}$A. At 4V, \emph{I}$\sim$10nA, indicating \textit{h} tunneling through hBN into 1L-WS$_2$. b) EL and PL positions from 4 different devices. The dashed line plots the unbiased PL position of X$^0$ measured in Fig.\ref{fig:Fig3}b}
\label{fig:Fig4}
\end{figure}
\begin{figure}
\centerline{\includegraphics[width=90mm]{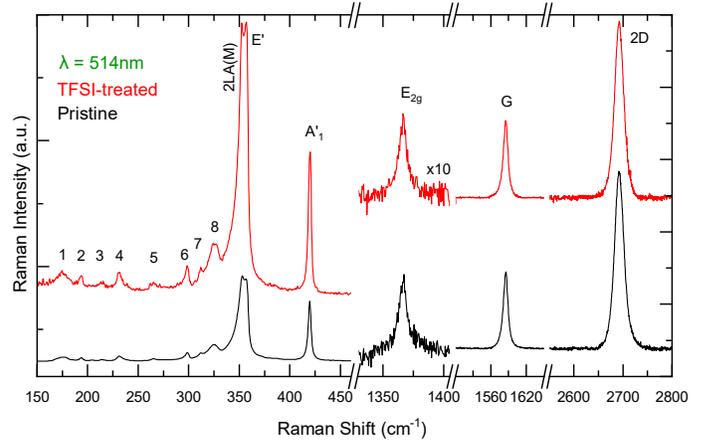}}
\caption{514.5nm Raman spectra of pristine (black line) and TFSI-treated (red line) 1L-WS$_2$/hBN/SLG LMH. The SLG and hBN Raman modes are labelled, as well as the modes for 1L-WS$_2$, as for Table 1. The 150-450cm$^{-1}$ (1300-2800cm$^{-1}$) ranges are normalized to the Si (2D) peaks, respectively. The E$_{2G}$ peak is multiplied by 10 for better visualization}
\label{fig:Fig5}
\end{figure}

To further understand the EL emission origin, we perform EL and PL spectroscopy at the same \emph{V}. Fig.\ref{fig:Fig4}a plots PL spectra at different \emph{V}. At \emph{V}=0V, the PL peak is$\sim$619.2nm ($\sim$2.002eV), assigned to X$^0$\cite{ZhuScRep2015,GutNL2013}. By increasing \emph{V} (i.e. increasing \textit{e} density in 1L-WS$_2$), a second peak appears at longer wavelengths ($\sim$630nm,$\sim$1.968eV), due to X$^{-}$\cite{ZhuScRep2015,ShangACSNano2015,ParAPL2017,PeimyooNL2014}. For \emph{V}$>$0V, the X$^0$ intensity gradually decreases and nearly vanishes, while X$^{-}$ shifts to longer wavelengths, Fig.\ref{fig:Fig4}a. This is expected for trionic emission, due to \textit{e}-doping induced by \textit{V}\cite{WangNL2017,YangNL2016,ZhuScRep2015,ShangACSNano2015,ParAPL2017,PeimyooNL2014}. Similar effects were observed in 1L-MoS$_2$/SiO$_2$/Si\cite{MakNatM2012}, hBN/1L-WSe$_2$/hBN/SiO$_2$/Si\cite{WangNL2019}, and hBN/1L-WS$_2$/hBN/SiO$_2$/Si\cite{KwakADM21}. Therefore, for similar tunneling current, EL agrees in energy and shape with the PL emission (see, e.g., the PL and EL spectra at the bottom of Fig.\ref{fig:Fig4}a). This is confirmed by Fig.\ref{fig:Fig4}b, where EL and PL peak positions are plotted for 4 devices, showing EL and PL emission at very similar wavelengths. Thus, EL predominantly originates from X$^{-}$\cite{CarmenNC2016,WangNL2017,ZhuScRep2015,ShangACSNano2015,WangNL2019}. The variations in X$^{-}$ energy for different LEDs are due to changes in charge carriers density across different samples. E.g., the charge density variation in 1L-WS$_2$ can be due to the number of vacancies in 1L-WS$_2$\cite{ZhangSci2014} and external impurities (PC residues and adsorbed water) after LED fabrication, which may vary from sample to sample.
\begin{table*}[htb]
\caption{Pos and (FWHM) in cm$^{-1}$ of WS$_2$ Raman peaks, before and after LMH assembly, and TFSI treatment}
\centering
\begin{tabular}{|c|c|c|c|c|c|c|}
\hline
Peak & Bulk-WS$_2$ Assignment & Bulk-WS$_2$  & 1L-WS$_2$ Assignment & 1L-WS$_2$-SiO$_2$ & 1L-WS$_2$-LMH & TFSI + 1L-WS$_2$-LMH\\ [0.5ex]
\hline
1 & LA(M) & 174.5 (11.1) & LA(M) & 175.6 (14.5) & 175.6 (14.6) & 174.9 (14.4) \\
2 & LA(K) & 194.8 (3.3) & LA(K) & 193.3 (4.5) & 193.8 (3.3) & 193.3 (4.7) \\
3 & A$_{1g}$(K)-LA(K) & 213.7 (4.2) & A$_{1}^{'}$(K)-LA(K) & 214.5 (5.7) & 214.5 (5.2) & 213.5 (6.0) \\
4 & A$_{1g}$(M)-LA(M) & 232.8 (5.7) & A$_{1}^{'}$(M)-LA(M) & 231.5 (6.7) & 231.9 (7.1) & 231.4 (5.9)\\
5 & A$_{1g}$(M)-ZA(M) & 266.8 (6.9) & A$_{1}^{'}$(M)-ZA(M) & 265.3 (6.9) & 265.9 (7.2) & 265.4 (7.0) \\
6 & E$^2_{2g}$($\Gamma$) & 297.6 (4.2) & E$^{''}$($\Gamma$) & 297.7 (2.8) & 298.5 (3.1) & 298.7 (2.6) \\
7 & LA(M)+TA(M) & 311.2(2.4) & LA(M)+TA(M) & 311.2 (2.5) & 311.8 (2.3) & 311.2 (2.4) \\
8 & E$^2_{2g}$(M) & 324.6 (17.5) & E$^{''}$(M) & 326.7 (25.5) & 325.9 (24.7) & 327.7 (25.7) \\
  & 2LA(M) & 350.6 (8.3) & 2LA(M) & 352.4 (9.3) & 352.7 (9.2) & 352.7 (8.0) \\
  & E$^{1}_{2g}$($\Gamma$) & 356.9 (1.5) & E$^{'}$($\Gamma$) & 357.2 (3.3) & 357.4 (3.1) & 357.2 (2.9) \\
  & A$_{1g}$($\Gamma$) & 420.8 (2.1) & A$_{1}^{'}$($\Gamma$) & 418.9 (3.9) & 419.8 (3.4) & 419.9 (3.4) \\ [1ex]
\hline %inserts single line
\end{tabular}
\end{table*}

We now consider the origin and consequences of excess \textit{e} in 1L-WS$_2$ for EL emission induced by \textit{V}. Besides the intrinsic charge carriers in 1L-WS$_2$ (typically \textit{n}-type due to S vacancies\cite{ZhangSci2014}), there is also an electrostatically induced charge in 1L-WS$_2$ when \emph{V}$>$0V. A SLG/hBN/1L-WS$_2$ tunneling junction acts as a MIS capacitor\cite{WangNL2017,WangNL2019,KwakADM21}. When \emph{V}$>$0 is applied to SLG, inducing positive charges in SLG, there is an opposite (negative) charge induced in 1L-WS$_2$\cite{WangNL2017,WangNL2019,KwakADM21}, thus making the charge density on 1L-WS$_2$ larger than for \emph{V}$=$0. When \emph{V}$>$\emph{V$_{ON}$}, \textit{h} will be injected by tunneling into 1L-WS$_2$ (Fig.\ref{fig:Fig1}d), hence, \textit{h} will recombine with \textit{e}. Consequently, the EL emission originates from X$^{-}$ states. However, the radiative recombination efficiency (defined as the number of \textit{e-h} pairs that recombine by emission of a photon divided by total number of \textit{e-h} pairs) of X$^{-}$ is lower than X$^{0}$ because of the small ($\sim$30meV) binding energy of trions\cite{JonesNatP2016}. Thus, to gain higher $\eta_{EL}$ one should favor X$^0$ EL emission by lowering the unbalanced free-carriers concentration in 1L-TMDs by either gate modulation\cite{WangNL2017,WangNL2019,KwakADM21,YangNL2016,SundaranNL2013,RossNN2014}, physical\cite{WangJPCL19,TongNL2013}, or chemical doping\cite{LiuNL2017,PeimyooNL2014}.

We thus treat 1L-WS$_2$ using TFSI to reduce doping and favor X$^0$ emission under bias and investigate the effects on EL emission and gated-PL. Fig.\ref{fig:Fig5} plots representative Raman spectra before (black) and after (red) TFSI treatment. By comparing the spectra before and after TFSI treatment, and the fits for the 1L-WS$_2$ in Table 1, we do not observe significant changes in peak position and FHWM. However, there is an overall intensity increase of the Raman modes of$\sim$50$\%$, compared to the Si peak. This indicates a reduction of \textit{n}-doping induced by TFSI treatment, because S vacancies in 1L-TMDs are commonly associated to \textit{n}-type behaviour and the reduction of these defects will reflect in \textit{p}-type doping fingerprint\cite{AmaniSci2015,AmaniACS2016,AmaniNL2016,DhakalJMCC2017}. Pos(A$^{'}_{1}$) is unaffected by TFSI treatment, which suggests that the reduction in the intrinsic 1L-WS$_2$ \textit{n}-doping induced by TFSI is$<<$10$^{12}$cm$^{-2}$\cite{SohierPRX2019}. Although TFSI is able to \textit{p}-dope SLG when it is in contact with the TFSI solution\cite{Heo2021}, Fig.\ref{fig:Fig5} shows negligible (within the errors\cite{Keftoot2022}) changes in the SLG (e.g. before (after): Pos(G)$\sim$1585.1 (1585.0)$\pm0.2$cm$^{-1}$, FWHM(G)$\sim$9.0 (9.1)$\pm0.2$cm$^{-1}$, Pos(2D)$\sim$2692.3 (2692.2)$\pm0.2$cm$^{-1}$, FWHM(2D)$\sim$20.9 (20.8)$\pm0.2$cm$^{-1}$, I(2D)/I(G)$\sim$2.4 (2.4), and A(2D)/A(G)$\sim$5.6 (5.6)) and hBN (e.g. before (after): Pos(E$_{2g}$)$\sim$1366.4 (1366.5)$\pm0.2$cm$^{-1}$ and FWHM(E$_{2g}$)$\sim$9.2 (9.1)$\pm0.2$cm$^{-1}$) Raman spectra after treatment, as both are protected by the top 1L-W$_2$.

\begin{figure}
\centerline{\includegraphics[width=90mm]{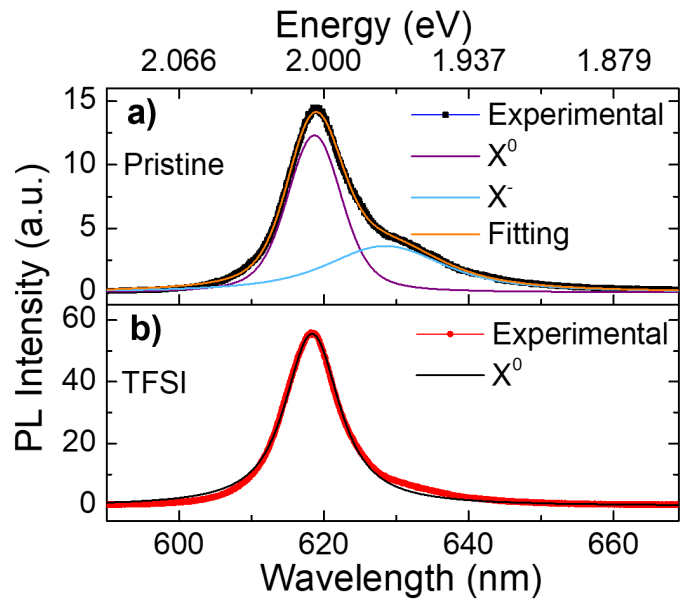}}
\caption{Fitting of PL spectra for (a) pristine and (b) TFSI-treated 1L-WS$_2$ on SiO$_2$/Si, for 532nm excitation}
\label{fig:Fig6}
\end{figure}

Fig.\ref{fig:Fig6}a plots a representative PL spectrum of 1L-WS$_2$ embedded in the LMH before TFSI treatment, and Fig.\ref{fig:Fig6}b after treatment. For the pristine case, there are two components, fitted by two Lorentzians$\sim$618.7nm ($\sim$2.004eV) and$\sim$629.1nm ($\sim$1.971eV) corresponding to X$^0$\cite{ZhaoNanoscale2013,GutNL2013} and X$^{-}$ emission\cite{ParAPL2017,PrandoPRA2021}. For non-biased devices, the spectral weight (defined as the area of each peak) of the PL emission indicates a majority emission due to X$^0$. After treatment, the PL emission evolves to a main single peak$\sim$618.1nm ($\sim$2.006eV), accompanied by a$\sim$4-fold increase in PL intensity. The changes in spectral weight of X$^0$ and X$^-$ emission after treatment can be assigned to a reduction in the \textit{e}-density in 1L-WS$_2$\cite{AmaniSci2015,AmaniACS2016,AmaniNL2016}, in agreement with our Raman analysis. Refs.\cite{TanohNL2019,BrestcherACS2021,Tanoh2021,AmaniSci2015,AmaniACS2016,AmaniNL2016,KimACS2016,MolasSRep2019,RoyNL2019,AlharbiAPL2017,CadizAPL2016,ZhangSRep2017,LinJMCC2019,GoodmanPRB2017,LuAPL2018} reported that PL enhancement depends on sample quality (defects) and may vary 1 to 10 times. In our samples we observe a PL increase$\sim$5$\pm$1-times, consistent with Refs.\cite{TanohNL2019,BrestcherACS2021,Tanoh2021,AmaniSci2015,AmaniACS2016,AmaniNL2016,KimACS2016,MolasSRep2019,RoyNL2019,AlharbiAPL2017,CadizAPL2016,ZhangSRep2017,LinJMCC2019,GoodmanPRB2017,LuAPL2018}.
\begin{figure}
\centerline{\includegraphics[width=90mm]{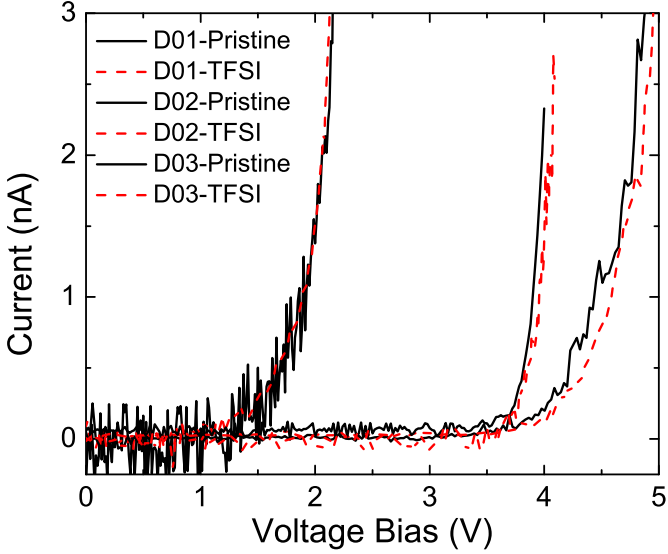}}
\caption{\emph{I}-\emph{V} curves of 3 LEDs before (solid black lines) and after (dashed red lines) TFSI treatment}
\label{fig:Fig7}
\end{figure}
\begin{figure}
\centerline{\includegraphics[width=80mm]{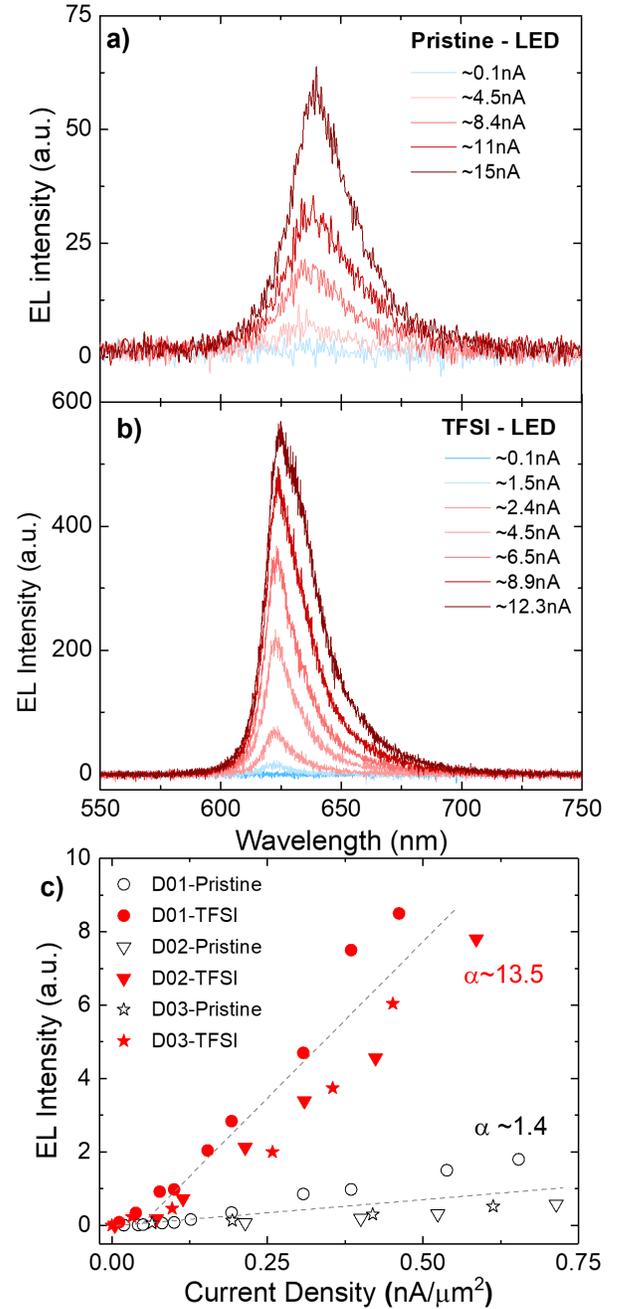}}
\caption{EL spectra from (a) pristine and (b) TFSI-treated 1L-WS$_2$-LEDs for different tunneling currents. AA$\sim$21$\mu$m$^2$. c) EL intensity as a function of tunneling current divided by AA for pristine (black) and TFSI-treated (red) 1L-WS$_2$-LEDs (3 devices). The dashed lines are a linear fit to the data}
\label{fig:Fig8}
\end{figure}

Fig.\ref{fig:Fig7}a plots typical \emph{I}-\emph{V} characteristics of 3 devices before (solid black lines) and after (dashed red lines) TFSI treatment. \emph{I} is not affected by the treatment. \emph{V}$_{ON}$ is mostly influenced by the hBN thickness\cite{HatACSN2015,Man2D2017}. Figs.\ref{fig:Fig8}a,b show EL collected before and after TFSI, respectively, for different \emph{I}. In both cases, EL is triggered for similar current levels (\emph{I}$<$5nA), and the intensity increases linearly with \emph{I}, Fig.\ref{fig:Fig8}c. The EL intensity slope as a function of current density (\emph{I} divided by AA) is affected by TFSI. For pristine-LEDs we get an average slope $\alpha\sim$1.4$\pm$0.3, while after TFSI $\alpha\sim$13.5$\pm$1.1, with 1 order of magnitude $\eta_{EL}$ increase, Fig.\ref{fig:Fig8}c. The red-shifts in the EL emission with \emph{I} increase in pristine ($<$6nm) and TFSI treated LEDs ($<$5nm), Figs.\ref{fig:Fig8}a,b, can be assigned to E$_F$ shift induced by the MIS structure\cite{SundaranNL2013,WangNL2019,AndrzejewskiACSPhot2019}.

Next, we estimate the external quantum efficiency (EQE) of our LEDs. This is defined as the ratio between the number of emitted photons (\emph{N}$_{ph}$) and that of injected \textit{h} per second (\emph{N}$_{h}$)\cite{SchubertBook2006}:
\begin{equation}
\label{eq:EQE}
EQE = \frac{N_{ph}}{N_{h}} = \frac{\sum_\mathbf{\lambda}N_{ph-counts}}{N_{h}}\times
\frac{A_{eff}}{\eta_{sys}},
\end{equation}
where $\sum_\mathbf{\lambda}N_{ph-counts}$ is the sum of the total photons collected by the spectrometer over the measured spectral range, A$_{eff}$=AA/A$_{spot}$, where A$_{spot}$ is the microscope objective spot size ({A$_{spot}$=$\pi [1.22\mathbf{\lambda}$/2NA}]$^2$ $\sim$2.2$\mu$m$^{2}$, with ${\lambda}$=618nm and NA=0.45), and $N_{h}$=\emph{I}$\times$\emph{t}/\emph{q}, where \emph{t} is the acquisition time, and \emph{q} the \textit{e} charge. The efficiency factor (defined as the ratio between the photons collected by the detector and the emitted photons by EL at the sample position) of our setup, including all optical components and spectrometer, is $\eta_{sys}\sim$0.0051, see Methods.

From Eq.\ref{eq:EQE} we get EQE$\sim$0.025\%$\pm$0.021\% and$\sim$0.195\%$\pm$0.324\% for pristine- and TFSI treated-LEDs, respectively, corresponding to a$\sim$8.7$\pm$1.5-fold increase, thus demonstrating that TFSI can boost EQE by almost one order of magnitude. It was reported that, using pulsed (AC) bias, EL emission can be enhanced a factor$\sim$4\cite{AndrzejewskiACSPhot2019} and up to$\sim$100 in a double optical cavity (distributed Bragg reflector (DBR) with an optical mirror)\cite{Zamudio2D2019}. Therefore, AC bias and photonic cavities could be combined with TFSI treatment to achieve EQE$>$10$\%$ in 1L-TMDs.
\begin{figure}
\centerline{\includegraphics[width=90mm]{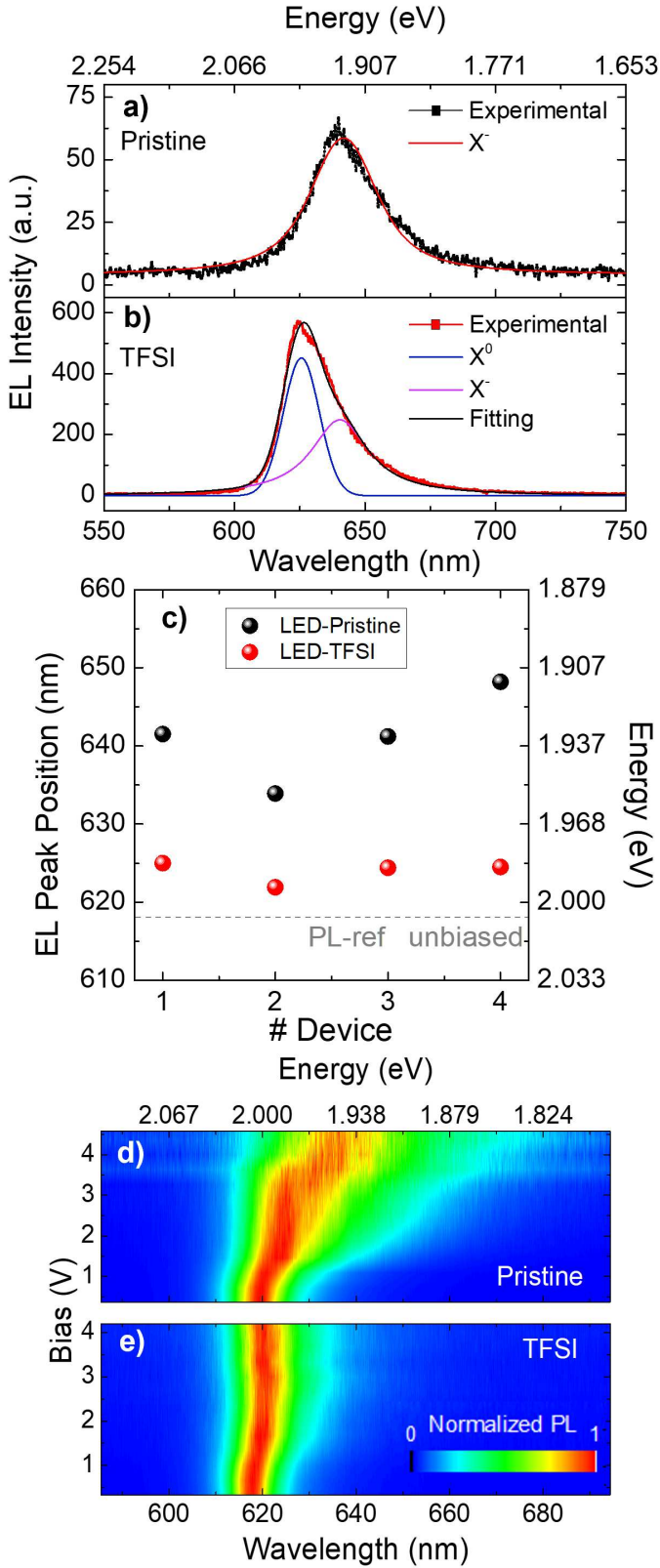}}
\caption{EL spectra of (a) pristine and (b)TFSI-treated LEDs at similar tunneling current$\sim$12nA, fitted with Lorentzians. c) Position of EL emission for different LEDs before (black) and after (red) TFSI. Color-plot of the gated-PL of (d) pristine and (e) TFSI-treated LED at similar laser excitation power and integration time}
\label{fig:Fig9}
\end{figure}

We now consider the EL emission features induced by TFSI treatment. By comparing EL before and after TFSI (Figs.\ref{fig:Fig8}a,b), a blue-shift in EL is observed. In pristine-LEDs, the EL emission is$\sim$641.8nm ($\sim$1.931eV), Fig.\ref{fig:Fig9}a, whereas after treatment it is$\sim$625.6nm ($\sim$1.982eV), Fig.\ref{fig:Fig9}b. Fig.\ref{fig:Fig9}c plots the EL peak position before and after treatment in 4 devices. After treatment, the EL emission shifts to shorter wavelengths, where X$^0$ is expected\cite{BerSRep2013,GutNL2013} (dashed line in Fig.\ref{fig:Fig9}c). In non-biased S-based TMDs devices, this shift could be due to the depletion of excess \textit{e} in \textit{n}-doped 1L-WS$_2$ due to TFSI\cite{AmaniSci2015,AmaniACS2016,AmaniNL2016,KimACS2016,MolasSRep2019,RoyNL2019,ZhangSRep2017,LinJMCC2019,GoodmanPRB2017}. Nevertheless, we cannot neglect the additional charge density induced by \emph{V} on the MIS capacitor. E.g. the \textit{I-V} characteristics in Fig.\ref{fig:Fig7} show that \emph{I} and \emph{V$_{ON}$} do not change before and after TFSI, suggesting the same tunneling condition is maintained across the 1L-WS$_2$/hBN/SLG junction. In both cases a comparable electric field (and electric charge) is developed across the junction for a given \emph{V}. Fig.\ref{fig:Fig7} implies that, independent of TFSI treatment, the same amount of negative charge is electrostatically induced in 1L-WS$_2$ at \emph{V}$>$0. However, taking into account the EL spectral shift towards X$^0$ emission upon bias, the expected depletion of excess \textit{e} in 1L-WS$_2$ cannot explain the electrical behaviour of Figs.\ref{fig:Fig9}b,c. Consequently, the emission profile is not compatible with the \emph{I-V} curves before and after TFSI in Fig.\ref{fig:Fig7}, given that the electric field across the junction should be modified by the \textit{e} density change in 1L-WS$_2$.

To get a better insight on the effects of TFSI on 1L-WS$_2$ based LEDs, Figs.\ref{fig:Fig9}d,e plot normalized PL spectra as a function of \emph{V} before and after TFSI. In the pristine case (Fig.\ref{fig:Fig9}d), the PL map shows an evolution in emission spectra from$\sim$620nm ($\sim$2.000eV) to$\sim$638nm ($\sim$1.943eV), corresponding to a spectral shift from X$^0$ to X$^{-}$ due to excess \textit{e} in 1L-WS$_2$ induced by \emph{V}. After TFSI treatment (Fig.\ref{fig:Fig9}e), the PL exhibits only a minor shift from$\sim$618nm ($\sim$2.006eV) to$\sim$622nm ($\sim$1.993eV), implying that the induced \textit{e}-charge in 1L-WS$_2$ does not contribute to the X$^{-}$ emission pathway. Therefore, similar to Figs.\ref{fig:Fig9}a,b, PL also indicates that the emission after TFSI treatment predominantly originates from radiative recombination of X$^0$, independent of \emph{V}. Refs.\cite{AmaniSci2015,AmaniNL2016,MolasSRep2019,RoyNL2019,AlharbiAPL2017,CadizAPL2016} claimed that TFSI treatment reduces the extent of \textit{n}-type behavior in S-based 1L-TMDs due to S vacancies passivation, consistent with the suppression of X$^{-}$ formation in Refs.\cite{KimACS2016,ZhangSRep2017,LinJMCC2019,GoodmanPRB2017,LuAPL2018}. Ref.\cite{LienSci2019} reported that TFSI acts as a Lewis acid, i.e. it can accept an \textit{e} pair from a donor\cite{Acid-def}, suppressing X$^{-}$ formation. Whereas Refs.\cite{TanohNL2019,BrestcherACS2021,Tanoh2021} claimed that TFSI may activate sub-gap states and reduce the \textit{n}-type behavior in S-based TMDs, as well as reducing X$^{-}$ formation. Our \textit{I-V}, EL and gated-PL results suggest that TFSI treatment i) depletes the excess \textit{e} in 1L-WS$_2$, acting as a Lewis acid\cite{LienSci2019} and ii) favours the radiative recombination of X$^{0}$ independent of bias, due to the activation of trapping states\cite{BrestcherACS2021,TanohNL2019} in 1L-WS$_2$ caused by the treatment. One would expect changes in the excitonic emission at such trapping states at RT, where the thermal energy can assist carrier de-trapping, and radiative recombination from excitons\cite{GoodmanPRB2017}. Therefore, the modification from non-radiative to radiative recombination by activation of trapping states could be further engineered to achieve more efficient optoelectronic devices.
\section{Conclusions}
We demonstrated a one order of magnitude enhancement in EL emission of 1L-WS$_2$-LEDs by performing TFSI treatment. EL predominantly originates from trions in pristine devices, while neutral excitons dominate in treated ones. The neutral excitonic emission is also restored in 1L-WS$_2$ gated-PL measurements. We attribute these changes to a reduction of \textit{n}-doping of 1L-WS$_2$, as well as changes in the relaxation and recombination pathways within 1L-WS$_2$. This paves the way to more efficient 1L-TMDs-based LEDs, and shed light into tunability of the excitonic emission of these devices.
\section{Methods}
\subsection{Raman characterization of LMH individual constituents}
Raman spectroscopy allows us to monitor LMs at every step of device fabrication. This should always be performed on individual LMs before and after assembly in LMHs and devices. This is an essential step to ensure reproducibility of the results, but, unfortunately, this is often neglected in literature.

Ultralow-frequency (ULF) Raman spectra in the range$\sim$10-50cm$^{-1}$ probe shear (C), corresponding to layer motion parallel to the planes, and layer breathing modes (LBM), corresponding to the motion perpendicular to them\cite{TanNM2012,ZhangPRB2013,Pizzi2020,FerNN2013}. Pos(C)$_N$ can be used to determine the number of layers\cite{TanNM2012,ZhangPRB2013,Pizzi2020} as N=$\pi({2\cos^{-1}[\frac{Pos(C)_{N}}{Pos(C)_\infty}]})^{-1}$, with Pos(C)$_\infty$ the bulk Pos(C).

\begin{figure}
\centerline{\includegraphics[width=90mm]{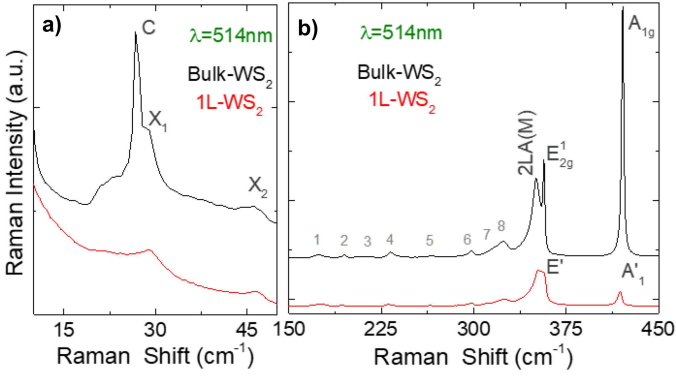}}
\caption{(a) Low- and (b) high-frequency 514.5nm Raman spectra of 1L-WS$_2$ (red) and bulk-WS$_2$ (black) on Si/SiO$_2$, normalized to the Si peak, with labels as for Table 1}
\label{fig:Fig10}
\end{figure}

Fig.\ref{fig:Fig10} plots the Raman spectra of non-treated 1L-WS$_2$ and bulk-WS$_2$. In Fig.\ref{fig:Fig10}a, the C mode and LBM are not observed for 1L-WS$_2$, as expected\cite{TanNM2012,ZhangPRB2013,Pizzi2020}. In bulk-WS$_2$, Pos(C)$\sim26.9\pm0.14$cm$^{-1}$. The spectral resolution$\pm0.14$cm$^{-1}$ for the ULF region is obtained as for Ref.\cite{Keftoot2022}. We observe two additional peaks$\sim$28.7$\pm0.14$cm$^{-1}$ and 46.4$\pm0.14$cm$^{-1}$, respectively, in agreement with Refs.\cite{Cheong2017,CheongNanoscale2015,Duesberg2015}. These do not depend on N\cite{Cheong2017,CheongNanoscale2015} and are seen because 514.5nm ($\sim$2.41eV) is nearly resonant with the B exciton ($\sim$2.4eV) of 1L-WS$_2$\cite{Shi2D2016,ZhaoACS2013,ZhuSR2014,CorroNL2016,ZhaoRS2020}, and$\sim$20meV above the bulk-WS$_2$ B exciton ($\sim$2.38eV)\cite{ZhaoACS2013,ZhuSR2014}. This gives rise to a resonant process\cite{Shi2D2016,ZhaoACS2013,ZhuSR2014,CorroNL2016,ZhaoRS2020}, which occurs because the laser energy matches the electronic transition of the B exciton, revealing features associated with intervalley scattering mediated by acoustic ph\cite{CarvalhoPRL2015,Carvalho2D2020,Gontijo2019}. A similar process also happens in 1L-MoS$_2$\cite{Cheong2017,CheongNanoscale2015} and other 1L-TMDs\cite{CarvalhoPRL2015,Carvalho2D2020,Gontijo2019}. Although our ULF filters cut$\sim$5cm$^{-1}$, the LBM is not detected in bulk-WS$_2$, as its frequency is expected to be$<$10cm$^{-1}$\cite{Pizzi2020}, because this resonant process with a 514.5nm laser reduces the signal to noise ratio in this spectral region\cite{Cheong2017}.

The high-frequency (HF) Raman spectra of non-treated 1L-WS$_2$ and bulk-WS$_2$ (Fig.\ref{fig:Fig10}b) show various peaks, Table 1. The first order Raman modes, i.e. E$^{'}$, A$^{'}_1$ in 1L-WS$_2$\cite{BerSRep2013,GutNL2013,Molina-SanchezPRB2011,ZhaoNanoscale2013} and E$^1_{2g}$, A$_{1g}$ in bulk-WS$_2$\cite{BerSRep2013,GutNL2013,Molina-SanchezPRB2011,ZhaoNanoscale2013}. E$^{'}$ (E$1_{2g}$) and A$^{'}_1$ (A$_{1g}$) correspond to in-plane and out-of-plane optical ph for 1L(bulk)-WS$_2$. Their nomenclature for 1L and bulk differs due to the different crystal symmetry\cite{BerSRep2013,GutNL2013,Molina-SanchezPRB2011,ZhaoNanoscale2013}. In 1L-WS$_2$ we get Pos(E$^{'}$)$\sim$356.8$\pm0.2$cm$^{-1}$, FWHM(E$^{'})\sim$3.2$\pm0.2$cm$^{-1}$, Pos(A$^{'}_1$)$\sim$418.5$\pm0.2$cm$^{-1}$, FWHM(A$^{'}_1$)$\sim$4.3$\pm0.2$cm$^{-1}$. In bulk-WS$_2$ we have Pos(E$^1_{2g}$)$\sim$356.8$\pm0.2$cm$^{-1}$, FWHM(E$^1_{2g}$)$\sim$1.5$\pm0.2$cm$^{-1}$, Pos(A$^{'}_1$)$\sim$420.8$\pm0.2$cm$^{-1}$, FWHM(A$^{'}_1$)$\sim$2.1$\pm0.2$cm$^{-1}$. In 1L-WS$_2$ the difference in peaks' position [Pos(E$^{'}$)-Pos(A$^{'}_1$)] is$\sim$61.7cm$^{-1}$ while this is$\sim$64.0cm$^{-1}$ in bulk-WS$_2$, further corroborating the identification of 1L\cite{BerSRep2013}. In the HF spectra of 1L- and bulk-WS$_2$ we also observe the 2LA(M) mode, involving two longitudinal acoustic (LA) ph close to the M point\cite{BerSRep2013,GutNL2013,Molina-SanchezPRB2011}. For 1L-WS$_2$ Pos(2LA(M))$\sim$351.9$\pm0.2$cm$^{-1}$ and FWHM(2LA(M))$\sim$9.2$\pm0.2$cm$^{-1}$, whereas for bulk-WS$_2$ Pos(2LA(M))$\sim$350.6$\pm0.2$cm$^{-1}$ and FWHM(2LA(M))$\sim$8.3$\pm0.2$cm$^{-1}$. The 2LA(M) mode originates from a second-order double resonant process\cite{CarvalhoPRL2015,Carvalho2D2020,Gontijo2019}, where momentum conservation is satisfied by two LA ph with opposite momenta around K- and M-points\cite{Carvalho2D2020}, therefore sensitive to differences in band structure between bulk and 1L-WS$_2$\cite{BerSRep2013,CarvalhoNatComm2017}.

I(A$_{1g}$)/I(E$_{1g}$)$\sim$3.2 in bulk-WS$_2$, where I is the peak height, is higher than I(A$^{'}_1$)/I(E$^{'}$)$\sim$0.8 in 1L-WS$_2$. I(2LA)/I(E$_{1g}$)$\sim$1 in bulk-WS$_2$ is lower than I(2LA(M))/I(E$^{'}$)$\sim$1.7 in 1L-WS$_2$. This can be explained considering that the main first-order (E$^{'}$, A$^{'}_1$) and second-order (2LA(M)) Raman modes are enhanced for 2.41eV excitation, due to exciton-ph coupling effects involving B exciton transitions\cite{CorroNL2016,CheiwchanchamnangijPRB2012}. These depend on mode symmetry (i.e. differ between out-of-plane and in-plane modes) as well as N\cite{CarvalhoPRL2015}. In bulk-WS$_2$, the out-of-plane A$_{1g}$ is resonant with the B exciton, unlike E$^1_{2g}$\cite{CarvalhoPRL2015}. The enhancement of A$_{1g}$ decreases with decreasing N due to the dependence of the lifetime of the intermediate excitonic states on N\cite{CarvalhoPRL2015}. The difference between I(2LA)/I(E$^{'}_1$) in 1L-WS$_2$ and I(2LA)/I(E$^1_{2g}$) in bulk-WS$_2$ is due to a change in band structure from direct bandgap in 1L to indirect in bulk-WS$_2$\cite{BerSRep2013,GutNL2013,Molina-SanchezPRB2011,ZhaoNanoscale2013}, which changes the double resonance conditions\cite{CarvalhoPRL2015,Carvalho2D2020,Gontijo2019}.

The Raman spectrum of 1L-WS$_2$ also shows 8 peaks in the range 170-350cm$^{-1}$ (Fig.\ref{fig:Fig10}b and Table 1). LA(M) and LA(K) correspond to one-ph processes originating from the LA branch at the M- and the K-points, respectively\cite{BerSRep2013,GutNL2013,Molina-SanchezPRB2011,ZhaoNanoscale2013}. Since LA(M) and LA(K) and E$^2_{2g}$(M) are one-ph processes from the edge of the BZ (q$\neq$0)\cite{BerSRep2013,GutNL2013,Molina-SanchezPRB2011,ZhaoNanoscale2013}, they should not be seen in the Raman spectra since, due to the Raman fundamental selection rule\cite{cardona}, one-ph processes are Raman active only for ph with q$\sim$0, whereas for multi-ph scattering the sum of ph momenta needs to be$\sim$0\cite{CarvalhoPRL2015,Carvalho2D2020,Gontijo2019,CarvalhoNatComm2017}. However these modes can be activated in presence of defects, as these can exchange momentum with ph, such that the sum of the momenta in the process is$\sim$0\cite{BerSRep2013,GutNL2013,Molina-SanchezPRB2011,ZhaoNanoscale2013}. A$_{1g}$(K)-LA(K), A$_{1g}$(M)-LA(M), A$_{1g}$(M)-ZA(M), LA(M)+TA(M) in bulk-WS$_2$ and A$^{'}$(K)-LA(K), A$^{'}_1$(M)-LA(M), A$^{'}_1$(M)-ZA(M), LA(M)+TA(M) in 1L-WS$_2$ are combinational modes, and Raman allowed\cite{BerSRep2013,GutNL2013,Molina-SanchezPRB2011,ZhaoNanoscale2013}. E$^2_{2g}$(M) correspond to a one-ph process originating from the transverse optical (TO) branch at the M-point\cite{BerSRep2013,GutNL2013,Molina-SanchezPRB2011,ZhaoNanoscale2013}. E$^2_{2g}$($\Gamma$) is a degenerate mode originating from the LO and TO branches at $\Gamma$\cite{BerSRep2013,GutNL2013,Molina-SanchezPRB2011,ZhaoNanoscale2013}.
\begin{figure}
\centerline{\includegraphics[width=90mm]{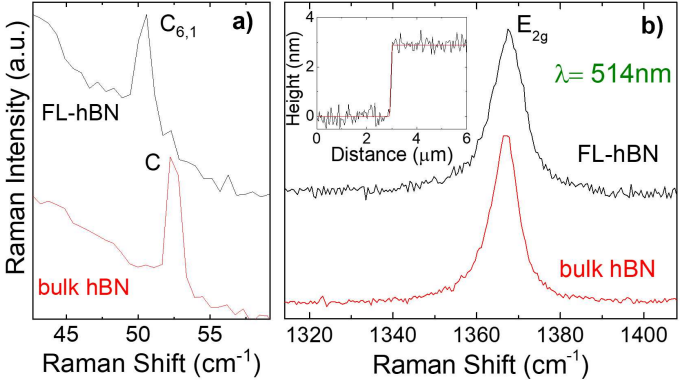}}
\caption{(a) ULF and (b) HF 514.5nm Raman spectra of$\sim$3nm hBN on Si/SiO$_2$ normalized to the Si peak. Inset: AFM height profile of the $\sim$3nm hBN on Si/SiO$_2$}
\label{fig:Fig11}
\end{figure}
\begin{figure}
\centerline{\includegraphics[width=90mm]{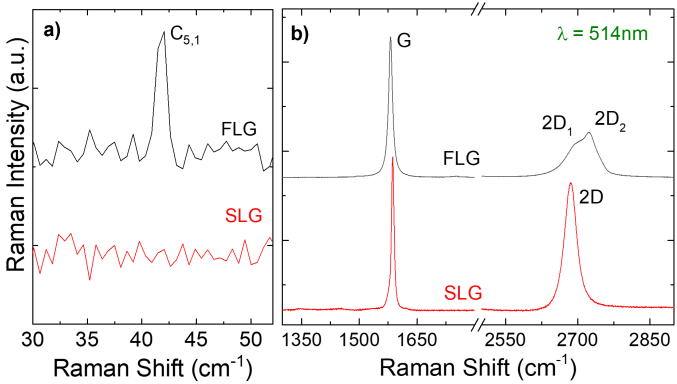}}
\caption{(a) ULF and (b) HF 514.5nm Raman spectra of SLG and FLG on Si/SiO$_2$ normalized to the Si peak}
\label{fig:Fig12}
\end{figure}

Fig.\ref{fig:Fig11} plots the Raman spectra of a$\sim$3nm hBN flake (black curves) and bulk-hBN (red curves). The latter has 2 Raman-active modes\cite{ReiPRB2005,AreNL2006}, C and E$_{2g}$. In Fig.\ref{fig:Fig11}a Pos(C)$_\infty$=52.3$\pm$0.14cm$^{-1}$ with FWHM$\sim$0.7$\pm0.2$cm$^{-1}$ for bulk-hBN and Pos(C)$_N$=50.4$\pm$0.14cm$^{-1}$ FWHM$\sim$0.8$\pm0.2$cm$^{-1}$ for the hBN flake. In bulk-hBN Pos(C)$_\infty$=$\frac{1}{\pi c}\sqrt{\frac{\alpha}{\mu}}$ =52.3cm$^{-1}$, with $\mu$ =6.9$\times$10$^{27}$kg$\AA^{-2}$  the mass of one layer per unit area, \textit{c} the speed of light in cm s$^{-1}$, and $\alpha$ the spring constant associated to the coupling between the adjacent layers\cite{Keftoot2022,Pizzi2020}. From this, we get $\alpha=16.9\times10^{18}$Nm$^{-3}$. From N=$\pi ({2\cos^{-1}[\frac{Pos(C)_{N}}{Pos(C)_\infty}]})^{-1}$, we get N=6$\pm$1 for the 3nm thick flake (measured with a Dimension Icon Bruker AFM in tapping mode) as shown in the inset of Fig.\ref{fig:Fig11}b). In Fig.\ref{fig:Fig11}b Pos(E$_{2g}$)$\sim$1368.5$\pm0.2$cm$^{-1}$ and FWHM(E$_{2g}$)$\sim$9.1$\pm0.2$cm$^{-1}$ for FL-hBN, and Pos(E$_{2g}$)$\sim$1367$\pm0.2$cm$^{-1}$ with FWHM(E$_{2g}$)$\sim$7.6$\pm0.2$cm$^{-1}$ for bulk-hBN. The peak broadening$\sim$1.5cm$^{-1}$ in FL-hBN can be attributed to strain variations within the laser spot, as thinner flakes conform more closely to the roughness of the underlying SiO$_2$\cite{Keftoot2022}. This is consistent with the fact that thicker hBN have lower root mean square (RMS) roughness\cite{Keftoot2022,Purdie2018,Cadore2D2022,Man2D2017}, e.g. 300nm SiO$_2$ has RMS roughness$\sim$1nm\cite{Man2D2017}, 2-8nm hBN has RMS roughness$\sim$0.2-0.6nm\cite{Keftoot2022}, while$>$10nm hBN thick presents RMS roughness$\sim$0.1nm\cite{Purdie2018,Man2D2017}.

The red curves in Figs.\ref{fig:Fig12}a,b are the Raman spectra of SLG on SiO$_2$/Si before LMH assembly. Pos(G)=1586.9$\pm0.2$cm$^{-1}$ with FWHM(G)=7.7$\pm0.2$cm$^{-1}$, Pos(2D)=$2685.2\pm$0.2cm$^{-1}$ with FWHM(2D)$\sim$29.3$\pm0.2$cm$^{-1}$, I(2D)/I(G)$\sim$0.85, A(2D)/A(G)$\sim$3.3. These indicate a \textit{p}-doping\cite{Basko2009,FerCom2007,FerNN2013} with E$_F\sim200\pm$50meV. No D peak is observed, thus negligible defects\cite{Cancado2011,FerCom2007,FerNN2013}. Pos(G) and Pos(2D) are affected by the presence strain\cite{FerCom2007,FerNN2013}. Biaxial strain can be differentiated from uniaxial from the absence of G-peak splitting with increasing $\epsilon$\cite{MohPRB2009,YoonPRL2011}, however at low ($\leq$0.5\%) $\epsilon$ the splitting cannot be resolved\cite{MohPRB2009,YoonPRL2011}. Thus, the presence (or coexistence) of biaxial strain cannot be ruled out. For uniaxial(biaxial) strain, Pos(G) shifts by $\Delta$Pos(G)/$\Delta\epsilon$ $\approx$23(60)cm$^{-1}$/\%\cite{MohPRB2009,YoonPRL2011}. Pos(G) also depends on doping\cite{Das2008,Basko2009}. E$_F\sim200\pm$50meV should correspond to Pos(G)$\sim$1584.3cm$^{-1}$ for unstrained SLG\cite{Das2008}. However, in our experiment Pos(G)$\sim$1586.9$\pm0.2$cm$^{-1}$, which implies a contribution from compressive uniaxial (biaxial) strain$\sim$0.1\% ($\sim$0.04\%). The black curves in Figs.\ref{fig:Fig12}a,b show the Raman spectrum of the FLG electrode on SiO$_2$/Si. Pos(G)$\sim1581.2\pm$0.2cm$^{-1}$ with FWHM$\sim$12$\pm0.2$cm$^{-1}$, Pos(2D$_1$)$\sim2694.0\pm$0.2cm$^{-1}$ with FWHM$\sim$48$\pm0.2$cm$^{-1}$, and Pos(2D$_2$)$\sim2725\pm$0.2cm$^{-1}$ with FWHM$\sim$33$\pm0.2$cm$^{-1}$. Pos(C)$_N\sim$41.4$\pm0.14$cm$^{-1}$, corresponding to N=5.
\subsection{Spectrometer efficiency}
The $\eta_{sys}$ of our spectrometer is derived as follows. We use a 50x objective (NA=0.45). Hence, the solid angle is $\theta$=(1-cos$\theta$)$\times$2$\pi$, where $\theta$=arcsin(NA/n), and n is the refractive index. Assuming n=1 we get $\theta$=0.672. Thus, M$_{50x-eff}$=$\theta$/(4$\pi$)$\times$100\%$\sim$5.4\%. In our Horiba system, the optical path from M$_{50x}$ to CCD includes 7 Mirrors (M$_{eff}\sim$83\%), a slit (S$_{eff}\sim$90\%), a grating (G$_{eff}\sim$60\%) and a CCD detector (CCD$_{eff}\sim$85\%). Therefore, the calculated overall collection+Horiba efficiency is: M$_{50x-eff}\times$(M$_{eff}$)$^7\times$S$_{eff}\times$G$_{eff}\times$CCD$_{eff}\sim$0.0067. To experimentally validate the calculation, we use a 0.5pW laser at 632.8nm and measure the counts at the CCD detector N$_{counts}$=149748. The photon energy at 632.8nm is E$_{ph}$=(1.24/0.638)$\times$1.6e$^{-19}$=3.13e$^{-19}$J. The laser power is P$_{opt}$=0.5e$^{-12}$ J/s. As a result, if the system efficiency is 100\% we expect to get 0.5e$^{-12}$/3.13e$^{-19}$=1597444 counts. Therefore, the Horiba system efficiency is Syst$_{eff}$=149748/1597444=0.094. Considering M$_{50x-eff}$, we get an overall collection + Horiba efficiency M$_{50x-eff}\times$Syst$_{eff}$=0.054$\times$0.094=0.0051, consistent with the theoretical estimation.
\section{Acknowledgments}
We acknowledge funding from the EU Graphene and Quantum Flagships, EU grant Graph-X, ERC Grants Hetero2D and GSYNCOR and GIPT, EPSRC Grants EP/K01711X/1, EP/K017144/1, EP/N010345/1, EP/L016087/1, EP/V000055/1, EP/X015742/1

\end{document}